\documentclass{aa}  
\usepackage{graphicx}
\usepackage{txfonts}
\begin{document}
   \title{The X-ray emission of the most luminous 3CR radio sources}


   \author{M. Salvati,
          \inst{1}
           G. Risaliti,
          \inst{1, 2}
           P. V\'eron 
          \inst{3}
          \and
           L. Woltjer
          \inst{1, 3}
          }

   \offprints{M. Salvati}

   \institute{INAF, Osservatorio di Arcetri,
              Largo E. Fermi 5, 50125 Firenze, Italy\\
              \email{salvati@arcetri.astro.it}
         \and
             Center for Astrophysics, Cambridge MA, USA\\
             \email{risaliti@arcetri.astro.it}
         \and
             Observatoire de Haute Provence, France\\
             \email{philippe.veron@oamp.fr}
             }

   \date{Received ; accepted }

 
  \abstract
   {Although many radio loud quasars and galaxies have been observed in 
    X~rays, systematic studies of well defined samples are rare.}
   {We investigate the X-ray properties of the most luminous radio sources
    in the 3CR catalogue, in order to assess if they are similar to the most
    luminous radio quiet quasars, for instance in the X-ray normalization
    with respect to the optical luminosity, or in the distribution of the
    absorption column density.}
   {We have selected the (optically identified) 3CR radio sources whose 
    178-MHz monochromatic luminosity lies in the highest factor-of-three bin.
    The 4 most luminous objects had already been observed in X~rays. Of the 
    remaining 16, we observed with XMM-Newton 8 randomly chosen ones, with the
    only requirement that half were of type~1 and half of type~2 according to
    the optical identification.}
   {All targets have been detected. The optical-to-Xray spectral index, 
    $\alpha_{ox}$, can be computed only for the type~1s and, in agreement
    with previous studies, is found to be flatter than in radio quiet quasars 
    of similar luminosity. However, the Compton thin type~2s have an absorption
    corrected X-ray luminosity syste\-ma\-ti\-cally lower than the type~1s, by
    a factor which makes them consistent with the radio quiet $\alpha_{ox}$.
    Within the limited statistics, the Compton thick objects seem to have a
    reflected component more luminous than the Compton thin ones.}
   {The extra X-ray component observed in type~1 radio loud quasars is beamed
    for intrinsic causes, and is not collimated by the absorbing torus as is
    the case for the (intrinsically isotropic) disk emission. The extra component
    can be associated with a relativistic outflow, provided that the flow opening 
    angle and the Doppler beaming factor are $\sim 1/5$ -- $1/7$ radians.}

   \keywords{galaxies: active --
                galaxies: quasars: general --
                radio continuum: galaxies --
                X-rays: galaxies }
   \authorrunning{M. Salvati et al.}
   \title{The X-ray emission of the most luminous 3CR radio sources}

   \maketitle
%

\section{Introduction}

A strong X-ray emission is a defining property of Active Galactic Nuclei,
and radio loud quasars and galaxies share this property. The most interesting
entries of the radio catalogues have been observed repeteadly with all the X-ray 
satellites launched so far. Although much has been learned, our knowledge is
still fragmentary in comparison with the radio quiet AGN. The latter
dominate the X-ray sky, and it is relatively straightforward to assemble
large samples which can then be studied at other wavelengths. On the contrary,
the luminous extragalactic radio sources are rare, and one has to observe
them one by one investing large amounts of satellite time. 

There are a few points on which a consensus has been established. The type~1 
radio loud quasars have an X-ray emission stronger than their radio quiet 
analogues of similar optical luminosity, which is quantified by a flatter $\alpha_{ox}$
(e.g. \cite{brink} versus \cite{steffen}). There are indications that the X-ray 
photon index $\Gamma$ is syste\-ma\-ti\-cally flatter in radio loud quasars 
(\cite{shastri}, but \cite {brink} have a more cautious view). At any rate, a flatter 
$\Gamma\sim 1.5$ -- 1 is well established in flat spectrum radio sources 
and low frequency peaked blazars (\cite{grandi}, \cite{fossati}). 
The type~2 radio galaxies are commonly found to host absorbed AGN,
a paradigmatic case is the near\-by powerful source Cygnus~A (\cite{young}). 
Even if absorbed below energies of several keV, the radio galaxies are
frequently detected around 1 keV at a level of a few percent of the main
emission (\cite{diana} and references therein). It is unclear if the residual 
emission is due to scattering of the main one, or is completely unrelated.

The basic picture is the extension of the so called unified scheme to the
case where the accretion disk is complemented with a relativistic outflow ("jet").
The radio galaxies and radio quasars are the misaligned and aligned members,
respectively, of the same population. The optical and soft X-ray emission of
the former is obscured by some intervening medium, perhaps arranged in
a toroidal geometry. The jet emission, because of the Doppler boosting,
becomes more and more prominent along a sequence where the line of sight gets 
closer and closer to the jet axis, passing from steep spectrum to flat spectrum 
radio quasars, and to blazars.

%
\begin{table*}
\caption{The sample objects.} 
\label{table:1}     
\centering         
\begin{tabular}{l c c c c c c}     
\hline\hline           
Name & Type & Redshift & Radio Luminosity & Magnitude & Exposure Time & References \\ 
& & $z$ & L$_{\rm r}$ & & (ks) & \\ 
\hline                     
  3C298 & Q & 1.439 & 29.79 & 16.79~(V) &  20~(C) & 1 \\ 
  3C9~~ & Q & 2.012 & 29.72 & 18.21~(V) &  16~(C) & 2 \\
  3C257 & G & 2.474 & 29.67 & 18.07~(K) &  30~(X) & 3 \\ 
  3C191 & Q & 1.956 & 29.55 & 18.65~(V) &  17~(C) & 4 \\
  3C239 & G & 1.781 & 29.46 & 22.50~(V) &  14~(X) & 5 \\
  3C454 & Q & 1.757 & 29.39 & 18.47~(V) &  16~(X) & 5 \\
  3C432 & Q & 1.785 & 29.38 & 17.96~(V) &  11~(X) & 5 \\
        &   &       &       &           &  20~(C) & 4 \\
  3C294 & G & 1.786 & 29.35 & 18.00~(K) & 118~(C) & 6 \\
  3C249 & G & 1.554 & 29.34 & 18.90~(K) &  35~(X) & 5 \\
  3C241 & G? & 1.617 & 29.30 & 23.50~(V) &  25~(X) & 5 \\
  3C318 & Q? & 1.574 & 29.30 & 20.30~(V) &  19~(X) & 5 \\
  3C205 & Q & 1.534 & 29.28 & 17.62~(V) &   5~(X) & 5 \\

\hline                  
\end{tabular}
\begin{list}{}{}
\item[] References. (1)~\cite{aneta}; (2)~\cite{fabian}; (3)~\cite{derry}; 
\item[] (4)~\cite{erlund}; (5)~our data; (6)~\cite{fabian2}. 
\end{list}
\end{table*}
%

In this paper we present and discuss a study of the X-ray properties of
the most luminous low frequency radio sources. By selecting at 178 MHz, the
frequency of the 3CR catalogue (\cite{3cr}), we select sources with a luminous
extended, isotropic component, which is thought to arise from the accumulation 
of relativistic plasma over the entire lifetime of the system. The AGN in our
sample are then expected to have had a large {\it time averaged} power in their past
life, whereas the X-ray lu\-mi\-no\-si\-ty depends on the {\it instantaneous} power at 
the present time. A correlation between the two holds only in a statistical sense.
On the other hand, by selecting with respect to an isotropic component we avoid 
all the uncertainties connected with the Doppler boosting. We restrict ourselves
to the 3CR sources at high Galactic latitude ($|\ell|>20^{\circ}$) with optical 
identification and redshift (\cite{ident}), and order them with respect to 
L$_{\rm r}$~, a radio luminosity parameter equal to the logarithm of the monochromatic 
luminosity at 178 MHz in W~Hz$^{-1}$~m$^{-2}$~s$^{-1}$. We adopt the concordance 
cosmology with $\Omega_{\rm m}=0.3, \Omega_{\Lambda}=0.7$, and $\rm H_{\circ}=
70~km~s^{-1}~Mpc^{-1}$. We neglect the $k$-correction because all the objects of 
interest happen to be steep spectrum radio sources, i.e. they have a
spectral slope $\alpha$ (f$_{\nu} \propto \nu^{-\alpha}$) close to 1 in the
100-MHz region. In half a decade, from L$_{\rm r}$~= 29.79 to L$_{\rm r}$~=29.28,
one has 20 objects.

The four most luminous ones had already been observed by previous investigators. We 
chose at random 8 additional objects among the remaining 16, with the only requirement 
that 4 were optical type~1 and 4 optical type~2, and observed them with XMM-Newton. 
One of the 8 (3C294) had previous Chandra observa\-tions, and we did not reobserve it. 
One further object (3C432) was also observed with Chandra after our XMM run had
been sche\-du\-led. Due to the randomness of our choice, we believe that the final
sample of 4+8 objects is a fair representation of the highest radio luminosity bin, 
although a rather slender one. The only infringement to randomness, i.e. the
preselection of optical types, implies that we cannot draw inferences about the
relative frequency of types from our X-ray data.

\section{Observations and data reduction}

In Table~\ref{table:1} we report the basic data of the objects in our sample.
The columns are: the 3C name; the optical classification, either a type~1 quasar (Q)
or a type~2 radio galaxy (G); the redshift $z$~; the radio luminosity parameter
L$_{\rm r}$~; the magnitude in the specified filter; the exposure time t$_{\rm exp}$
in kiloseconds, and the satellite used for the X-ray observation (either X or C for 
XMM-Newton and Chandra, respectively); and, finally, the reference to the X-ray data. The
optical data are taken from \cite{ident} and from the NASA/IPAC Extragalactic Database 
(NED); in case of discrepancy, we take the most recent determination; if no data are 
available in the above catalogues, we resort to the X-ray references listed in the Table.

A question mark appended to the optical type indicates an intermediate 
classification, with characteristic properties of both a quasar and a galaxy present 
in the same object. In particular, the galaxy 3C241 has been found to 
exhibit a broad line in the near-infrared observer-frame spectrum (\cite{hirst}), 
and the broad line object 3C318 has a very red, underluminous optical continuum 
(\cite{willott}).  
For the sake of future discussion, it must be stressed again that all the objects in 
Table~\ref{table:1} are steep spectrum radio sources; this is usually interpreted as 
evidence of viewing angles substantially larger than the jet beaming.

The data have been processed according to the standard sequence with version
7.0.0 of the XMM SAS package (http://\-xmm.\-esac.\-esa.\-int/\-sas/). Since we were
interested in the nucleus only, we extracted the source events in an aperture 
of the order of the PSF width (typically three times); the background events
were extracted far from possible extended features (jets and lobes); if a
diffuse cluster emission was also present, the background extraction region was
an annulus immediately exterior to the source.

The analysis software was version 11.3 of XSPEC 
(http://\-heasarc.\-gsfc.\-nasa.\-gov/\-docs/\-xanadu/\-xspec/). 
If there were a sufficient number of counts, 
the spectra were rebinned to a minimum of 15 counts per bin, and analyzed with the 
$\chi^2$ statistic; otherwise no rebinning was made, and the C statistic was used. 
The relatively low signal to noise ratio prompted the adoption of very simple
spectral models: the most complex combination was

\begin{center}
    wabs*(powerlaw+zgauss+zwabs*powerlaw) \\
\end{center}

\noindent with the same photon index for the two power laws, and the local
absorption fixed at the Galactic value. Quite often even simpler models proved 
adequate. Our assumption on the photon indices implies the (conservative)
assumption that the low energy component is due to hot reflection or to partial 
covering.

%
\begin{table*}
\caption{Results of the X-ray analysis.}
\label{table:2}
\centering
\begin{tabular}{l l c c c c c } 
\hline\hline
Name  & Type & Intrinsic Luminosity & Spectral Index &  Column Density &Equivalent Width & Reflected Luminosity \\
& & L$_{2-10}$ (10$^{45}$ erg s$^{-1}$) & $\Gamma$ & N$_{\rm H}$ (10$^{22}$ cm$^{-2}$) & (keV)  &L$_{\rm refl}$ (10$^{44}$ erg s$^{-1}$)  \\
\hline
3C298 &QU&~~~~ 24.7 ~~~~&~~~~1.82 ($+0.03, -0.03$)~~~~&~~~~0.45 ($+0.03, -0.03$)~~~~&~~~...~~&~~~...~~\\
3C9 &QU&~~~~ 2.92 ~~~~&~~~~1.58 ($+0.10, -0.10$)~~~~&~~~...~~~~~&~~~~...~~~~&~~~~...~~~~\\
3C257 &GA&~~~~ 0.64 ~~~~&~~~~1.42 ($+0.30, -0.21$)~~~~&~~~~5.5 ($+6.4, -2.7$)~~~~&~~~~...~~~~&~~~~$<$0.64~~~~\\
3C191 &QU&~ 3.93 ($+0.54, -0.09$)~&~~~~1.73 ($+0.28, -0.12$)~~~~&~~~~0.38 ($+0.26, -0.28$)~~~~&~~~...~~~~~&~~~~...~~~~\\
3C239 &GT&~...~&~~~2.10 ($+0.65, -0.45$)~~~&~~~$< 0.32$~~~&~~~2.5 ($+2.5, -1.8$)~~~&~2.2 ($+0.3, -0.3$)~\\
3C454 &QU&~ 3.30 ($+0.15, -0.15$)~&~~~1.68 ($+0.09, -0.07$)~~~&~~~$< 0.13$~~~&~~~$< 0.17$~~~&~~~...~~~\\
3C432 &QU&~ 2.47 ($+0.16, -0.16$)~&~~~1.91 ($+0.20, -0.14$)~~~&~~~$< 0.32$~~~&~~~$< 0.66$~~~&~~~...~~~\\
3C294 &GA&~ 0.80 ($+0.16, -0.16$)~&~1.9~&~84 ($+11, -9$)~&~0.20 ($+0.13, -0.11$)~&~0.19 ($+0.24, -0.08$)~\\
3C249 &GT&~~~~~...~~~~~~~&~1.76 ($+0.32, -0.18$)~&~~~$< 0.82$~~~&~~~$< 0.95$~~~&~2.3 ($+0.4, -0.4$)~\\
3C241 &G?A&~ 1.27 ($+0.21, -0.17$)~&~1.98 ($+0.32, -0.30$)~&~13 ($+5, -4$)~&~0.49 ($+0.33, -0.31$)~&~1.03 ($+0.48, -0.42$)~\\
3C318 &Q?U&~ 1.76 ($+0.08, -0.08$)~&~1.95 ($+0.14, -0.13$)~&~0.29 ($+0.17, -0.16$)~&~$< 0.34$~&~~~...~~~~\\
3C205 &QU&~ 7.09 ($+0.33, -0.33$)~&~2.07 ($+0.15, -0.11$)~&~0.43 ($+0.18, -0.17$)~&~$< 0.26$~&~~~...~~~~\\
\hline
\end{tabular}
\end{table*}
%

The results are presented in Table~\ref{table:2}, which contains: the 3C 
name; the optical and the X-ray type, where the latter is coded U for unabsorbed 
(N$_{\rm H} < 10^{22}$ cm$^{-2}$), A for absorbed (10$^{22}$ cm$^{-2} < $ 
N$_{\rm H} < 10^{24}$ cm$^{-2}$), and T for Compton thick (10$^{24}$ cm$^{-2} < 
$ N$_{\rm H}$); the absorption corrected luminosity in the rest frame 2--10 keV 
interval, L$_{2-10}$ (this component cannot be measured in Compton thick sources); 
the photon index $\Gamma$; the column density N$_{\rm H}$ intrinsic to the 
source; the equivalent width EW of an Fe line, computed in the rest frame with 
respect to the absorption corrected continuum; the reflected luminosity in the rest 
frame 2--10 keV interval, L$_{\rm refl}$ (this component cannot be measured if 
there is a negligible intrinsic N$_{\rm H}$). All errors or upper limits are at 
the 90\% confidence level for one interesting parameter. Sometimes, the data retrieved 
from the literature are incomplete.

A point to be noted immediately is the close correspondence between the optical
and the X-ray types: all quasars are unabsorbed, while all galaxies (with the
possible exception of 3C249) are absorbed, either Compton thin or thick. 
A clearcut Compton thick example is 3C239: here only the reflected component is 
visible, with a spectral index similar to the unabsorbed sources and a very low
upper limit on the intrinsic N$_{\rm H}$. The true origin of this component
(reflection instead of underluminous unabsorbed emission) is shown 
by the large equivalent width of the Fe line. The line energy 
is not compatible with neutral iron [6.67 (+0.17-0.13) keV], suggesting 
that at least in this case the visible component is indeed reflected from a hot 
medium which preserves the intrinsic spectral index; the width of the line
is very weakly constrained (~$< 0.5$ keV).

   \begin{figure}[h!]
      \resizebox{\hsize}{!}{\includegraphics[angle=-90]{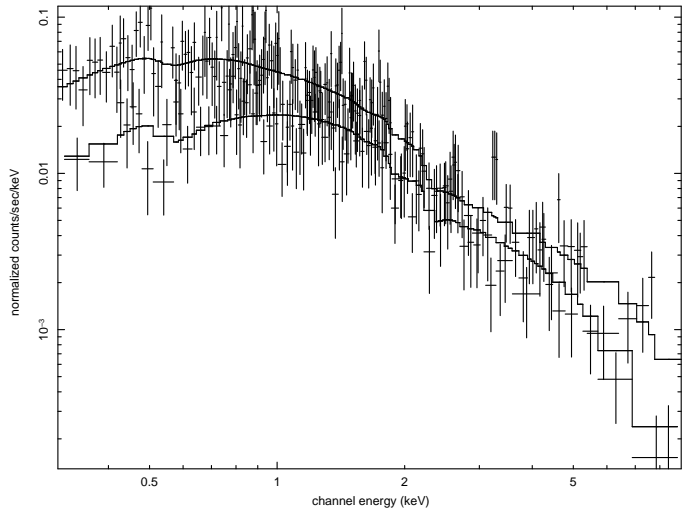}}
      \caption{Data and fitted model for 3C454. The upper and lower sets refer
        to PN and MOS, respectively.}
      \label{figure:1}
   \end{figure}
%

   \begin{figure}[h!]
      \resizebox{\hsize}{!}{\includegraphics[angle=-90]{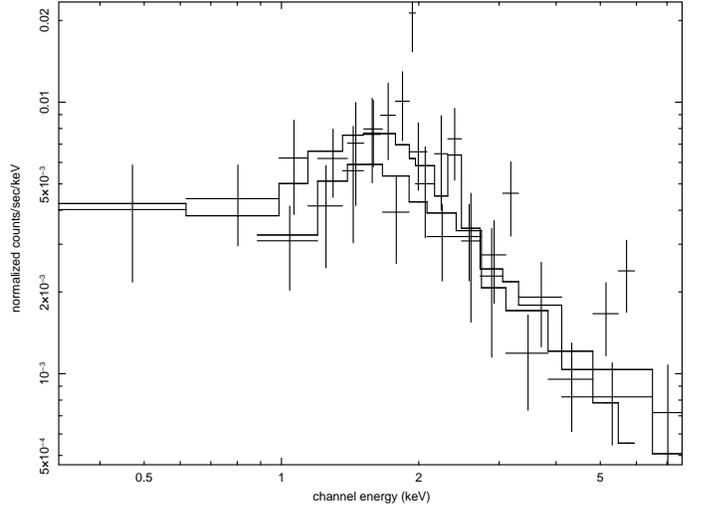}}
      \caption{Same as Fig.~\ref{figure:1} for 3C241.}
      \label{figure:2}
   \end{figure}
%

   \begin{figure}[h!]
      \resizebox{\hsize}{!}{\includegraphics[angle=-90]{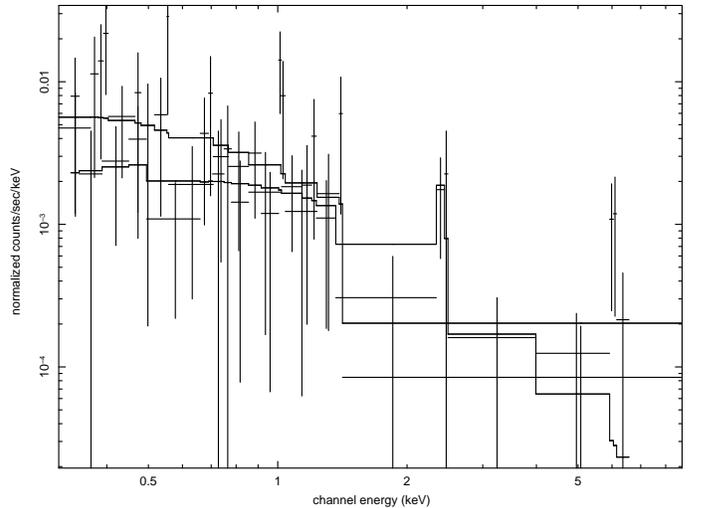}}
      \caption{Same as Fig.~\ref{figure:1} for 3C239.}
      \label{figure:3}
   \end{figure}
%

The X-ray spectrum of 3C249 is similar to 3C239, but here the Fe line is not
detected. We classify it as a Compton thick source on the basis of the optical type 
and the relative weakness of the X-ray emission, however one should keep in mind
that this particular object is compatible with our findings but does not add
independent evidence to them. The point representing 3C249 in Fig.~\ref{figure:4} is 
bracketed between parentheses in order to signal the uncertain X-ray classification.

%
\begin{table*}
\caption{The additional sample.}
\label{table:3}
\centering
\begin{tabular}{l c c c c c c}
\hline\hline
Name & Type & Redshift & Radio Luminosity & Intrinsic Luminosity & Reflected Luminosity & References \\
& & $z$ & L$_{\rm r}$ & Log(L$_{2-10}$) & Log(L$_{\rm refl}$) & \\
\hline
  3C324   & GT & 1.206 & 29.12 & ... & 43.66 & 1 \\
  3C380   & QU & 0.691 & 29.10 & 45.81 & ... & 2 \\
  3C280   & GT & 0.997 & 29.09 & ... & 43.84  & 3 \\
  3C325   & G?A & 1.135 & 29.05 & 44.85 & $<$43.45 & 1 \\
  3C309.1 & QU & 0.904 & 28.97 & 45.78 & ... & 2 \\
  3C212   & QU & 1.049 & 28.95 & 45.78 & ... & 4 \\
  3C210   & GT & 1.169 & 28.87 & ... & 43.85 & 1 \\
  3C184   & GA & 0.994 & 28.83 & 44.76 & 43.14 & 2 \\
  3C295   & GA & 0.461 & 28.82 & 44.48 & 42.50 & 5 \\
  3C265   & GA & 0.811 & 28.79 & 44.70 & 43.40 & 6 \\
  3C254   & QU & 0.734 & 28.69 & 45.32 & ... & 2 \\

\hline
\end{tabular}
\begin{list}{}{}
\item[] References. (1)~our analysis of Chandra archival data; (2)~\cite{diana}; (3)~\cite{dona}; 
\item[] (4)~\cite{ald}; (5)~\cite{hard}; (6)~\cite{bond}.
\end{list}
\end{table*}
%

Also to be noted is the segregation in X-ray luminosity between type~1s and Compton
thin type~2s: the former are more luminous than 2.5 10$^{45}$ erg s$^{-1}$, while 
the latter (after correction for absorption) are less luminous than 
8.0 10$^{44}$ erg s$^{-1}$; the logarithmic mean is 45.70$\pm$0.37 
and 44.85$\pm$0.07, respectively. The two objects with intermediate optical 
classification, 3C318 and 3C241, fall in between the two groups with 
45.22$\pm$0.03. A consequence of the systematic difference of intrinsic 
luminosity, which is enhanced by the effects of absorption, is that the 
Fe line is detected only in type~2 sources. The equivalent widths with respect to 
the intrinsic continuum are analogous to the radio quiet AGN (e.g. \cite{ew}). In all 
cases (except 3C239) the line energy is compatible with neutral iron; for consistency, 
also the upper limits to the line EW in type~1s are computed at 6.4 keV.

Figures \ref{figure:1}, \ref{figure:2}, and \ref{figure:3} show the X-ray spectra 
typical of unabsorbed, Compton thin, and Compton thick sources (3C454, 3C241, and 
3C239, respectively).

\section{Discussion}

The main point of the present work has been already anticipated in the previous
Section, and is the segregation in intrinsic luminosity between the unabsorbed
type~1s and the absorption corrected, Compton thin type~2s. This will be shown to
have far reaching implications. However our finding is plagued by the small
number of objects involved. In order to put the result on firmer grounds, we
have collected all the 3CR sources with X-ray observations having: optical
identification and redshift; high Galactic latitude ($|\ell|>20^{\circ}$);
and radio luminosity parameter L$_{\rm r}$ in the factor-of-three 
interval immediately below our fiducial sample, i.e. $ 28.67 < \rm L_{\rm r}
< 29.28$.

The additional sample contains 46 objects, of which only 12 observed in
X~rays. With respect to the fiducial sample, the additional one is sparsely 
observed (26\% versus 60\%); furthermore, we have no control on the criteria 
by which the observed sources were chosen. In order to confirm or disprove 
the luminosity segregation which we want to investigate, a crucial point is
a possible bias of the additional sample with respect to the orientation
of the jet axis. One source is the well known blazar 3C454.3; its X-ray 
emission is completely dominated by the jet, so we do not consider it further 
in the following. All the remaining 11 sources have steep radio spectra,
and none of them is classified as a blazar; for the sake of completeness,
we point out that the spectra of three of them (3C380, 3C309.1 and 3C212) 
become flat above a few GHz, perhaps a hint that their orientation angles 
are at the lower end of the non-blazar region. Also the distribution 
of optical properties (4 type 1's, 6 type 2's, and one intermediate type) 
does not exhibit obvious peculiarities.

Table~\ref{table:3} gives some basic information about the 11 additional sources;
more specifically we list: the 3C name; the optical and X-ray type; the
redshift $z$; the radio luminosity parameter L$_{\rm r}$; the intrinsic and reflected
X-ray luminosity, L$_{2-10}$ and L$_{\rm refl}$ (actually, the logarithm of 
the luminosity in erg s$^{-1}$); and the reference to the X-ray data. 
Note that 3C325 is a further case of intermediate optical classification (\cite{grimes}).

   \begin{figure}[h!]
      \resizebox{\hsize}{!}{\includegraphics[angle=0]{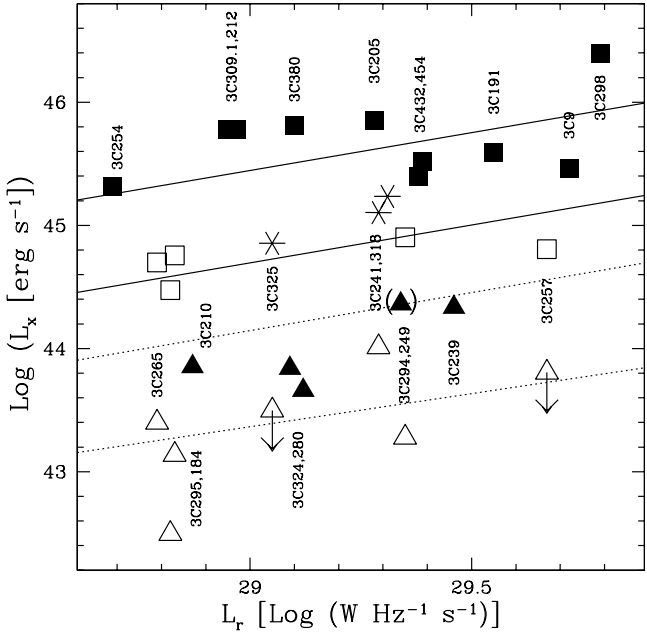}}
      \caption{The X-ray luminosity of the sources in our sample (fiducial plus 
       additional), plotted against the radio luminosity parameter. Filled squares, 
       quasars; stars, intermediate optical types; empty squares, Compton thin 
       galaxies (the absorption corrected direct component); filled triangles, 
       Compton thick galaxies; filled triangle between parentheses, 3C249 (see
       text); empty triangles, Compton thin galaxies (the reflected component).
       The upper and lower solid lines sketch the regions occupied by type 1's
       and type 2's, respectively; the dotted lines represent 
       the 5\% fraction of the solid lines.}
      \label{figure:4}
   \end{figure}
%

Figure \ref{figure:4} summarizes the information of Tables \ref{table:2} and 
\ref{table:3}: we plot the logarithm of the X-ray luminosity, Log(L$_{2-10}$)
and Log(L$_{\rm refl}$), against the radio luminosity parameter, L$_{\rm r}$. 
The meaning of the symbols is explained in the caption.
For the sake of clarity we omit the errorbars: the radio fluxes have typical 
errors of 1 Jy at the 3C flux limit of 10 Jy, so that the horizontal errorbars
are less than 0.04 dex; the X-ray values have typical errors of less than 25\%, 
i.e. less than 0.1 dex. Only the reflected components of Compton thin galaxies, 
the empty triangles in the Figure, have larger errors of the order of 0.3 dex. 

The lower right corner is expected to be underpopulated, given the limit of 10~Jy
for the 3C catalogue and a typical depth of 10$^{-15}$ erg cm$^{-2}$ s$^{-1}$ for 
the X-ray exposures.

The two parallel solid lines sketch the regions occupied by the type~1s and the
Compton thin type~2s, respectively. The difference of about 0.75 dex is fully 
consistent with the average values found in the previous Section for the fiducial 
sample only, i.e. the addition of more sources confirms the luminosity segregation
of the two types. Also, the intermediate optical types are confirmed to lie at 
intermediate X-ray luminosities. The dotted lines indicate the 5\% fraction of
the type~1 and type~2 luminosity, respectively. This particular value is generally 
regarded as an upper limit for the reflected or scattered components, because of 
various constraints on geometry, efficiency, and optical depth. 

In the case of type~1s the optical emission of the AGN is visible, and one can
compute the index $\alpha_{ox}$. We do that for all the type~1s of the fiducial
sample; we derive the rest frame monochromatic fluxes at 2 keV and at 2500 \AA\ 
by means of the observed $\Gamma$ in the X~rays, and by assuming $\alpha$ = 0.5
in the optical (f$_{\nu} \propto \nu^{-\alpha}$). We find $\alpha_{ox} = 1.35 
\pm 0.11$. At the average optical flux of our sources, 1.9 $10^{31}$ erg cm$^{-2}$ 
s$^{-1}$ Hz$^{-1}$, the radio quiet AGN have $\alpha_{ox} = 1.65 \pm 0.05$
(\cite{steffen}), thus we recover the well known result (cf. the Introduction)
that radio loud quasars are more X-ray luminous than radio quiet ones of the
same optical luminosity. 

This is not true for the radio galaxies, however. If we take into account the
lower intrinsic X-ray luminosity of our type~2s we find $\alpha_{ox} \sim 1.64$,
fully compatible with the radio quiet one. Here we do not observe directly the
optical emission of the AGN, and we have assumed that type~2s and type~1s with
the same L$_{\rm r}$ have the same intrinsic optical luminosity. The additional
component characteristic of the jetted sources becomes visible at smaller and
smaller viewing angles for longer and longer wavelengths; indeed, even the high
luminosity blazars exhibit optical broad lines with equivalent widths within a
factor of two of the non blazar AGN (\cite{pian}, \cite{wills}). This points to the 
jet optical emission never being dominant over the disk optical emission, and gives
some ground to our assumption.

The X-ray normalization relative to the optical, and the (meager) evidence 
provided by the X-ray spectral features (slope $\Gamma$, equivalent width of
the Fe line), suggest that at the large viewing angles typical of radio
galaxies we see only the X-ray emission due to an accretion disk very similar
to the radio quiet AGN. The self-consistent normalization of X-ray, optical
and low frequency radio emission does not leave much room for "fossil" quasars, 
where the extended radio lobes survive for a substantial time after the central 
engine is switched off.

The additional X-ray emission connected with the ``radio 
loudness'' is not seen in radio galaxies, not because of the intervening 
absorption, but because this emission is intrinsically beamed.
The obvious explanation for the beaming is the relativistic aberration of
the jet emission. One can try a more quantitative assessment by adopting the
distribution of viewing angles proposed by Barthel (\cite{bart}), according
to whom radio loud AGN appear as type~1s or type~2s if their viewing angle
is smaller or larger than $\sim 44^{\circ}$; thus the average viewing angle
is $\theta_1\sim 31^{\circ}$ and $\theta_2\sim 69^{\circ}$ for the two classes, 
respectively. Within the type~1 class, one finds the blazar subclass which
corresponds to viewing angles smaller than the jet beaming angle, $\rm \theta_b < 
10^{\circ}$.

If one chooses the most favorable scenario, i.e. a jet of constant length
and a very flat spectral slope ($\Gamma=1$), the ratio of the brightness
in the two directions $\rm \theta_b$ and $\theta_1$ is

$$ \rm R_X = [(1-\beta cos\theta_1)/(1-\beta cos\theta_b)]^2,  \qquad
   \beta = \sqrt{1-\gamma^{-2}}                       $$

\noindent where $\gamma$ is the bulk Lorentz factor of the jet. This ratio is
equal to about 80 for $\gamma = 15$, as suggested by the analysis of the wide
band spectral energy distribution of blazars (\cite{ghisa}). Indeed, by using 
the maximum viewing angle for a blazar, $\rm \theta_b$, instead of the average
one, we have further underestimated $\rm R_X$. We conclude that the typical X-ray 
luminosity of high L$_{\rm r}$ blazars should be $ \rm R_X$ times the typical
X-ray luminosity of steep spectrum quasars of comparable L$_{\rm r}$, i.e. at 
least 3 10$^{47}$ erg s$^{-1}$. No such blazar is known all over the sky. 
A possible solution to this inconsistency could be a geometrically wide, stratified 
jet: here the narrow and fast spine would be responsible for the blazar properties, 
while the slower (but still relativistic) sheath would produce the wide angle 
additional component. Similar schemes have been suggested already, both for AGN 
(\cite{chiabe}) and for Gamma Ray Bursts (\cite{grb}).

The above equation for $\rm R_X$ can in principle be applied to the two
directions $\theta_1$ and $\theta_2$ to derive the residual jet emission in
radio galaxies. This however has nothing to do with the L$_{\rm refl}$ points
in Fig.~\ref{figure:4}: they refer to a spectral component produced above the 
absorbing material, at a distance of $\sim$parsecs from the source, whereas 
time variability constrains the blazar emission to much smaller dimensions
(e.g. \cite{ghisa}).

The correct comparison for the L$_{\rm refl}$ points is with the 5-percent-fraction
dotted lines. One notes that in Fig.~\ref{figure:4} the empty triangles
lie around or below the lower dotted line, which implies that the reflected
component of Compton thin type~2s has the expected normalization with respect to
the intrinsic disk component. The filled triangles, instead, are systematically
higher, and lie around or below the higher dotted line, as if they were normalized
to the wide jet component. In other words, if one insisted in attributing the 
visible emission of Compton thick objects to reprocessing of the disk emission 
only, one would have to accept reprocessed fractions as high as 30\%.

The statistics is certainly not compelling, however a possible scenario is one
where the optical depth at large viewing angles is not the same in all objects.
Compton thin and Compton thick type~2 objects would be viewed at similar angles,
around $70^{\circ}$, and would have relatively little and relatively large amounts
of matter around them, respectively. Thus the Compton thickness would
be related with a larger covered solid angle, perhaps so large as to reprocess
also the moderately beamed wide jet emission. Under this hypothesis the peculiar
position of 3C241 is nicely accounted for. This source exhibits at the same time
an intermediate X-ray luminosity and a relatively large N$_{\rm H}$, as if the
line of sight were grazing at the same time the boundaries of the wide jet and 
the boundaries of a particularly large absorber: indeed, here the reprocessing 
fraction is as high as 8\%, and the corresponding empty triangle falls right in 
the region of Compton thick sources.

\section{Summary}

We have studied the X-ray properties of a sample representative of the
most luminous 3CR sources. There is a clear difference in absorption corrected
X-ray luminosity between the type~1s and the type~2s in our sample, with the
latter being a factor of about 6 less luminous than the former. The difference
does not change if we augment our sample with similar objects retrieved from 
the literature.

While the optical--to--X-ray slope $\alpha_{ox}$ of the type~1s is flatter,
the type~2s have the same $\alpha_{ox}$ of the radio quiet AGN of similar
power.

We argue that, at least at high radio luminosities, the quasi-isotropic,
disk-produced X-ray emission is at the same level as in radio quiet AGN, and
that the additional emission related with the ``radio loudness'' is anisotropic
for intrinsic reasons, not connected with the circumnuclear absorption. The
anisotropy is likely due to Doppler beaming, however there are arguments which
disfavor a single zone narrow jet with a Lorentz factor $\sim$~15: perhaps
these values apply only to the spine of the jet, with a wider and slower
sheath being responsible for the additional emission of type~1s.

We find marginal evidence that Compton thick sources have larger amounts
of absorbing matter around them, which extends so close to the axis of the
system as to be able to reprocess the emission of the wide jet as well as 
the disk.

\end{document}